\UseRawInputEncoding 

\documentclass[twocolumn,showpacs,floats,floatfix,superscriptaddress,aps,prl]{revtex4-1}

\usepackage{amssymb}
\usepackage{graphicx}
\usepackage{amsmath}
\usepackage[colorlinks=true,linkcolor=blue,citecolor=blue,urlcolor=blue]{hyperref}

\def\rev{\textcolor{black}}

\begin{document}

\title{Self-healing of non-Hermitian topological skin modes}

\author{Stefano Longhi}
\thanks{stefano.longhi@polimi.it}
\affiliation{Dipartimento di Fisica, Politecnico di Milano, Piazza L. da Vinci 32, I-20133 Milano, Italy}
\affiliation{IFISC (UIB-CSIC), Instituto de Fisica Interdisciplinar y Sistemas Complejos - Palma de Mallorca, Spain}
\date{\today}

\begin{abstract}
A unique feature of non-Hermitian (NH) systems is the NH skin effect, i.e. the edge localization of an extensive number of bulk-band eigenstates in a lattice with open or semi-infinite boundaries. Unlike extended Bloch waves in Hermitian systems, the skin modes are normalizable eigenstates of the Hamiltonian that originate from the intrinsic non-Hermitian point-gap topology of the Bloch band energy spectra. Here we unravel a fascinating property of NH skin modes, namely self-healing, i.e. the ability to self-reconstruct  their shape after being scattered off by a space-time potential. 
 \end{abstract}

\maketitle
{\em Introduction.} Self-healing is the fantastic property of certain classical or quantum (matter) waves  to reconstruct their original shape after being scattered off by a potential (an obstacle) \cite{A1,A2,A3}. Such a special property is rather generally shared by diffraction-free and thus non-normalizable (delocalized) states of the underlying wave equation. Important examples include Bessel waves of the Helmholtz equation \cite{A1,A2,A4} and self-accelerating (Airy) waves of the Schr\"odinger equation \cite{A3,A5,A6}. Self-healing has been demonstrated for optical \cite{A1,A3,A8,A8a,A8b,A8c}, acoustic \cite{A9,A10,A11,A12} and matter waves \cite{A13,A14}, with a variety of applications in different areas of science such as in microscopy and biomedical imaging \cite{A15,A16,A17}, material processing \cite{A18}, particle manipulation \cite{A19,A20}, sensing \cite{A8a,A8b,A8c} and quantum communications \cite{A21}. However, in a norm-preserving (Hermitian) system any normalizable (bound) wave function cannot be strictly self-healing. An interesting and open question is whether {\em infinitely-many} self-healing {\em normalizable} waves can exist  in NH systems \cite{r1}. An important class of such systems is provided by NH lattices, where the role of topology and its far-reaching physical consequences are attracting an enormous interest \cite{r1a,r2,r3,r4,r5,r6,r7,r8,r9,r10,r11,r12,r13,r14,r15,r16,r17,r18,r19,r20,r20a,r21,r22,r23,r24,r25,r26,r27,r28,r28a,r29,r30,r31,r32,r33,r34,r34a,r34b,r35,r36,r37,r38,r39,r40,r41,r42,r43,r44,r45,r46,r47,r48,r49,r50,r51,r52,r53,r54,r55,r56,r57,r58,r59,r60,r61,r62,r63,r64,r65,r66,r67,r68,r69,r70,Referee1,Referee2}
 (for a recent review see \cite{r51}). A unique feature of NH lattices is the skin effect \cite{r5,r6,r7,r9,r29}, i.e. the localization of an extensive number of bulk eigenstates at the edges under open (OBC) or semi-infinite (SIBC) boundary conditions. 
 The localized skin modes replace the extended Bloch waves of Hermitian lattices and their origin can be traced back to the nontrivial point-gap topology of the bulk energy spectra under periodic boundary conditions (PBC), thus establishing a bulk-edge correspondence for skin modes \cite{r3,r29}.\\ 
In this work we unveil that topological skin edge modes share the fascinating property of being self-healing waves. Like non-normalizable diffraction-free waves in Hermitian systems, in one-dimensional (1D) NH lattices with SIBC there are infinitely many localized (normalizable) topological skin edge states that can reconstruct their shape after being scattered off by a rather arbitrary space-time potential.\\
\\
\textit{Wave self-healing.}  Let us consider the time-dependent dynamics of a wave function $|\psi(t) \rangle$ described by the Schr\"odinger-like wave equation
\begin{equation}
i \frac{d}{dt} | \psi \rangle = (\hat{H}+\hat{V} ) | \psi \rangle \label{eq1}
\end{equation}
where $\hat{H}$ is the time-independent Hamiltonian of the system, which is assumed rather generally NH, and $\hat{V}=\hat{V}(t)$ describes a space-time local scattering potential (the $^{\prime}$obstacle$^{\prime}$), which vanishes for $t>T$ and with compact support in space [Fig.1(a)]. At initial time $t=0$ the system is prepared in the state $| \psi(0) \rangle= |\phi(0) \rangle$, and let $| \phi(t) \rangle$ be the evolved wave function in the absence of the scattering potential $\hat{V}$, i.e. $| \phi(t) \rangle= \exp(-i \hat{H}t) | \phi (0) \rangle$. 
Clearly, the presence of the scattering potential destroys the unperturbed evolution of the wave function, so that after interaction with the potential, i.e. for $t>T$, $|\psi(t) \rangle$ can largely deviate for ever from the unperturbed solution $|\phi(t) \rangle$. The wave function $|\phi(t) \rangle$ is dubbed self-healing \rev{ if the deviation $\ | \xi(t) \rangle \equiv |\psi(t) \rangle- | \phi(t) \rangle$ is asymptotically much smaller than $ | \phi(t) \rangle$ as $t \rightarrow \infty$} regardless of the form of $\hat{V}$, i.e. provided that [Fig.1(a)] $\lim_{t \rightarrow \infty} \epsilon(t)=0$, where
\begin{equation}
\epsilon(t) = \frac{ \langle \xi(t) |  \xi (t) \rangle}{\langle \phi(t) | \phi(t) \rangle}. \label{eq2}
\end{equation}
 \rev {Note that the above condition corresponds to $ \| \tilde{\psi}(t)-\tilde{\phi}(t) \| \rightarrow 0$ for the normalized wave functions $ | \tilde{\psi}(t) \rangle= | \psi(t) \rangle / \| \psi(t) \|$ and $ | \tilde{\phi}(t) \rangle= | \phi(t) \rangle / \| \phi(t) \|$.}
Clearly, in an Hermitian system owing to norm conservation any normalizable wave function is not strictly self-healing, though it can approximate an extended (non-normalizable) wave function at some extent \cite{A6}. For example, for a freely-moving quantum particle in a one-dimensional space, $\hat{H}=-\partial^2 / \partial x^2$, the self-accelerating Airy solutions to the time-dependent Schr\"odinger equation
 \cite{A5} are non-normalizable self-healing waves \cite{A3}. Other non-normalizable self-healing modes include Bessel waves, parabolic cylinder waves,
Weber and Mathieu beams, Bloch surface waves, and others (see e.g. \cite{A2,A8b,Morandotti}). However, in a NH system propagation-invariant normalizable waves can be found \cite{Yuce}.\\
\begin{figure}[t]
  \centering
    \includegraphics[width=0.45\textwidth]{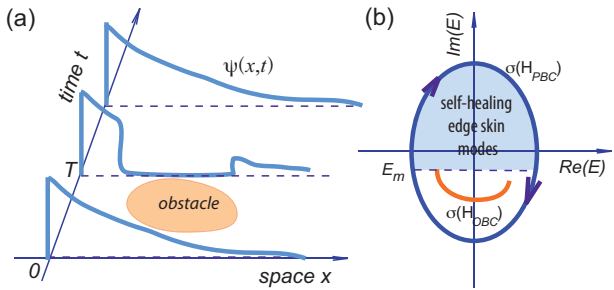}
   \caption{(a) Sketch of wave function propagation and self-healing property. After being scattered off by a space-time localized potential (the $^{\prime}$obstable$^{\prime}$), the wave function ${\psi}(x,t)$ can reconstruct its shape, as if the scattering potential were not present. (b) In a NH semi-infinite lattice with a left boundary, any topological edge skin mode at energy $E$ with $W(E)<0$ and ${\rm Im}(E)>E_m$  (shaded area in the figure) is a self-healing wave function. In the figure, the outer closed loop describes the energy spectrum $\sigma(H_{PBC})$, whereas the inner open arc is the energy spectrum $\sigma(H_{OBC})$.}
     \label{fig1}
\end{figure}

\textit{Energy spectra, topological skin modes and the bulk-edge correspondence.}
We consider a one-dimensional NH lattice with short-range hopping with Hamiltonian $\hat{H}$ in physical space given by 
\begin{equation}
\hat{H}=\sum_{n,l=1}^N H_{n,l} |n \rangle \langle l|, 
\end{equation}
where $H_{n,l}$ is a $N \times N$ banded matrix and  $N$ is the number of lattice sites. We indicate by $H_{PBC}$ and $H_{OBC}$ the $N \times N$ matrix Hamiltonians under PBC and OBC, respectively, in the large (thermodynamic) $N$ limit. For a single-band model,  $H_{OBC}$ is a banded Toeplitz matrix, i.e. $(H_{OBC})_{n,l}=t_{n-l}$ with $t_{n}=0$ for $n>s$ and $n<-r$ ($t_{-r}, t_s \neq 0$), where $t_{\pm l}$ are the left/right hopping amplitudes among sites distant $\pm l$ in the lattice and $r,s \geq 1$ are the largest orders of left/right hopping. $H_{PBC}$ is a circulant matrix with the same form as $H_{OBC}$, except for the top right and bottom left corners of the matrix. Finally, we indicate by $H_{SIBC}$ the infinite-dimensional matrix Hamiltonian under SIBC with a boundary on the left but not on the right, i.e. $(H_{SIBC})_{n,l}=t_{n-l}$ for $n,l=1,2,3,...$. The central result in the band theory of NH systems is that the energy spectra $\sigma (H_{PBC})$, $\sigma (H_{OBC})$ and $\sigma (H_{SIBC})$ are rather generally distinct, which implies the emergence of the NH skin effect, topological NH edge states and the need for a non-Bloch band theory. These results, studied in several recent works \cite{
r6,r14,r16,r29,r33,r34} and briefly reviewed in Sec.1 of \cite{Supplemental}, are basically rooted in the spectral theory of non-self-adjoint Toeplitz matrices and operators \cite{S1,S2,S3,S4}. Specifically, for a single-band lattice: (i) $\sigma(H_{PBC})$ is a closed loop in complex energy plane described by the Bloch Hamiltonian $H(k)=P(\beta= \exp(i k))$, where $P(\beta)=\sum_{l=-r}^{s}t_l \beta^l$ is the Laurent polynomial associated to the Toeplitz matrix and $-\pi \leq k < \pi$ is the Bloch wave number. (ii) $\sigma(H_{OBC})$ is the set of complex energies $E=P( \beta)$, where $\beta$ varies on the generalized Brillouin zone (GBZ) $C_{\beta}$. $\sigma(H_{OBC})$ is always topological trivial in terms of a point gap \cite{r29}. The definition and calculation of $C_{\beta}$ is discussed in \cite{r6,r14,r33,r34}, and briefly reviewed in \cite{Supplemental}. (iii) $\sigma(H_{SIBC})=\sigma(H_{PBC}) \bigcup B $, where $B$ is the interior of the PBC energy spectrum loop such that for $E\in B $ the winding number $W(E)$, defined by
\begin{equation}
W(E)= \frac{1}{2 \pi i} \int_{-\pi}^{\pi} dk  \frac{d}{dk} \log  \det \left\{ H(k)-E \right\}
\end{equation} 
is non vanishing. If $W(E)<0$, then $E$ is an eigenvalue of $H_{SIBC}$ of multiplicity $|W(E)|$, and the corresponding (right) eigenvectors are exponentially localized at the left edge. 
Such a result provides a bulk-boundary correspondence for NH systems, relating the appearance of skin edge states in a semi-infinite lattice to the topology of the PBC energy spectrum \cite{r29}.\\
 
 \textit{Self-healing of topological skin modes.} The central result of this work is that in NH lattices displaying the NH skin effect there are infinitely many skin edge modes that are self-healing. Specifically, let us consider a one-dimensional NH lattice with SIBC, with a boundary on the left but no boundary on the right, and with a GBZ $C_{\beta}$ that is, at least partly, external to the unit circle (to ensure the existence of left-edge skin states). The  local scattering potential is assumed to have a compact support both in space and time, i.e. $\hat{V}= V_n(t) | n \rangle \langle n |$ with $V_n(t)=0$ for $t>T$ and $n>L$. Let us indicate by $E_{m_1}$  the largest imaginary part of the energies in the set $\sigma(H_{OBC})$, i.e. $E_{m1}={\rm max}_{\beta \in C_{\beta}} {\rm Im} \{ P(\beta) \}$;  $E_{m_2}$ the largest imaginary part of the energies $E$ in the set $B$ defined by $\{E \in B \; | \; W(E)>0  \; \}$; and $E_m={\rm max}(E_{m_1}, E_{m_2})$. Note that the set $B$ is empty if the GBZ is entirely external to the unit circle $|\beta|=1$, i.e. if there are not Bloch point\ \cite{r14}; in this case one should assume $E_m=E_{m1}$ [as in Fig.1(b)].
 The following {\em{theorem}} can be then proven, which is illustrated in Fig.1: any topological skin edge state $|\phi(t) \rangle=|\phi_0 \rangle \exp(-i E_0 t)$ with energy $E_0$ \rev{ and $W(E_0)<0$ is self healing if and only if ${\rm Im}(E_0) > E_m$}.\\
  \rev{A simple corollary of this theorem is that any topological skin edge state belonging to $H_{OBC}$ is not self-healing, because in this case one has ${\rm Im}(E_0) \leq E_{m1} \leq E_m$.}\\
 Here we provide a sketch of the proof of the theorem (technical details are given in \cite{Supplemental}). Let us indicate by $|\psi(t) \rangle$ the wave function satisfying Eq.(\ref{eq1}) with the initial condition $| \psi(0) \rangle= | \phi_0 \rangle$, and let $| \xi(t) \rangle = | \psi(t) \rangle - | \phi(t) \rangle$ be the deviation of the wave function $| \psi(t) \rangle$ from the unperturbed (skin edge eigenstate) solution. The proof consists of two main steps. In the first step, one shows that, after interaction with the scatting potential, the deviation $\xi_n(T)= \langle n | \xi(T) \rangle$ vanishes as $ n \rightarrow \infty$ faster than exponential, i.e. for any $h>0$ one has $\lim_{n \rightarrow \infty} \xi_n(T) \exp(h n) =0$. Physically, this result stems from the fact that, since the hopping in the lattice is finite (short range) and the scattering potential has a limited support in space ($V_n=0$ for $n>L$), the speed of excitation spreading in the lattice arising from the interaction with the scattering potential is bounded (according to the Lieb-Robinson bound \cite{r3}), and thus after interaction  $\xi_n(T)$ remains basically unperturbed, i.e. very close to zero, for large enough $n$. The fast decay of $\xi_n$ with $n$ is mathematically justified by the asymptotic form of the exponential of a banded matrix \cite{S5} (Sec.2 of \cite{Supplemental}). Let us then indicate by $| \beta \rangle$ the set of eigenfunctions of $H_{OBC}$ (skin modes) with energy $P(\beta)$ belonging to $ \sigma (H_{OBC})$, i.e. $H_{OBC} | \beta \rangle=P(\beta) | \beta \rangle$ with $\beta \in C_{\beta}$.  Note that $| \beta \rangle$ is also an eigenstate of $H_{SIBC}$ when $|\beta|>1$ in the $N \rightarrow \infty$ limit. For large $n$,  $ \langle n | \beta \rangle$ behaves as $ \langle n | \beta \rangle \sim \beta^{-n}(1+ A_{\beta} \exp(- i \theta_{\beta} n) )$ with some $\beta$-dependent constants $A_{\beta}$ and $\theta_{\beta}$. Since $| \xi (T) \rangle$ is bounded with a localization higher than any exponential, one can decompose $| \xi (T) \rangle$ as a superposition (integral) of $| \beta \rangle$ skin states, i.e. one can write (Sec.1 of \cite{Supplemental})
 \begin{figure}[t]
  \centering
    \includegraphics[width=0.45\textwidth]{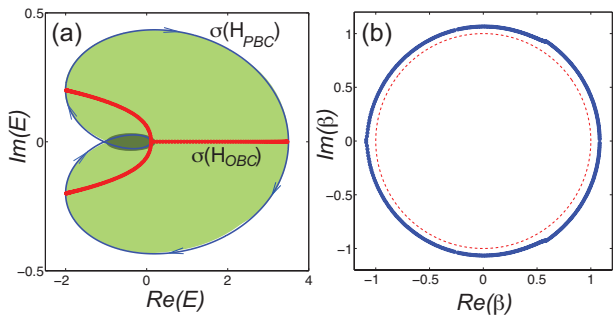}
   \caption{(a) Energy spectrum of $H_{PBC}$ (outer thin closed loop with one self-intersection), $H_{OBC}$ (inner bold open arcs) and $H_{SIBC}$ (shaded areas) of a NH lattice with nearest- and next-to-nearest neighbor hopping amplitudes 
   $t_{-2}=1$, $t_{-1}=1$, $t_0=0$, $t_1=0.7$, and $t_2=0.8$. In the light shaded area $W(E)=-1$, corresponding to simple (non-degenerate) skin edge state, whereas in the dark shaded area $W(E)=-2$, corresponding to the existence of two energy-degenerate skin edge states of $H_{SIBC}$. The largest value $E_{m_1}$ of ${\rm Im} (\sigma(H_{OBC}))$ is $E_{m_1}=0.2$. (b) The numerically-computed GBZ $C_{\beta}$, corresponding to a deformed circle with $|\beta|>1$ all along $C_{\beta}$. The thin dashed curve depicts the reference unit circle $|\beta|=1$.}
     \label{fig2}
\end{figure}
$ | \xi(T) \rangle= \oint_{C_{\beta}} d \beta F( \beta) | \beta \rangle$
with $F( \beta)$ non-singular on $C_{\beta}$. Since $\hat{V}=0$ for $t>T$, after the scattering event the wave function $ | \xi(t) \rangle$ evolves according to the Schr\"odinger equation $i \partial_t | \xi \rangle= \hat{H}_{SIBC} | \xi \rangle$, so that for $t>T$ one has
$ | \xi(t) \rangle= \oint_{C_{\beta}} d \beta F( \beta) \exp [-i P(\beta) (t-T)] \;  | \beta \rangle.$
The second step is to calculate the growth rate of $\|  \xi(t) \|^2=\langle \xi(t) | \xi(t) \rangle$. To this aim, one has to distinguish two cases (Sec.3 of \cite{Supplemental}). If $C_{\beta}$ is entirely external to the unit circle, i.e. $|\beta|>1$ for any $\beta \in C_{\beta}$, the growth rate of $\|  \xi(t) \|$  is $E_{m1}={\rm max}_{\beta \in C_{\beta}} {\rm Im} (P(\beta))$, which is attained at the value $\beta_s \in C_{\beta}$ corresponding to the most unstable saddle point of $P(\beta)$. Since $ \| \phi(t) \| $ grows in time as $\sim \exp ( {\rm Im}(E_0) t)$, one has $	\lim_{t \rightarrow \infty} \epsilon(t)=0$ \rev{if and only if } ${\rm Im}(E_0)> E_m$, where $\epsilon(t)$ is defined by Eq.(\ref{eq2}) and $E_{m}=E_{m_1}$. On the other hand, if a portion of $C_{\beta}$ is internal to the unit circle \rev{ the asymptotic analysis shows that the growth rate of $\|  \xi(t) \|$  is the larger number between $E_{m_1}$ and $E_{m_2}$}, where $E_{m_2}$ is the largest imaginary part of energies in the set $B$ \cite{Supplemental}. This proves the theorem. $\blacksquare$ \\

 As an illustrative example, let us consider a lattice with nearest- and next-nearest neighbor hopping ($r=s=2$). Figure 2 shows the energy spectra $\sigma(H_{PBC})$, $\sigma(H_{OBC})$ and $\sigma(H_{SIBC})$ and corresponding GBZ, which is entirely external to the unit circle with $E_{m}=E_{m_1} \simeq 0.5$. In the wide light shaded region of Fig.2(a), for each complex energy $E$ there is a single topological skin edge state ($W=-1$), while when $E$ is internal to the narrow dark shaded region encircling the origin there are two linearly-independent skin edge states ($W=-2$). To show the self-healing property of skin edge states,  we consider a strongly absorbing potential $V_n(t)=-10 i$ which is non-vanishing in the interval $2<t<4$ and in the spatial region $1 \leq n \leq L=10$.  The initial state $ | \phi_0 \rangle$ is chosen to be a skin edge state with an energy $E_0$ in the stable (${\rm Im}(E_0)>E_m$) or unstable (${\rm Im}(E_0)<E_m$) regions.  The self-healing property is measured by the long-time behavior of $\epsilon(t)$ [Eq.(\ref{eq2})]. Figure 3 illustrates the typical numerical results of wave propagation in the lattice, corresponding to the self-healing of the skin mode for ${\rm Im}(E_0)>E_m$   [Fig.3(a)], and to the disruption of the skin mode for ${\rm Im}(E_0)<E_m$ [Fig.3(b)]. \rev{The results are obtained by solving Eq.(\ref{eq1}) in Wannier (real-space) basis by an accurate fourth-order Runge-Kutta method on a  finite-sized lattice with OBC and with a size wide enough ($N=300$ sites) to avoid right-edge effects over the largest propagation  time ($ t \sim 20$), which would destroy the SIBC skin state \cite{r3,rLonghi21}. A strategic method to selectively prepare the system in a self-healing SIBC edge state is discussed in \cite{rLonghi21} and in Sec.5 of \cite{Supplemental}. } As clearly shown in the left panel of Fig.3(a), the strongly absorbing potential cuts the excitation at lattice sites $n \leq L$, however after the scattering process the skin edge state can restore its original shape, corresponding to a vanishing of $\epsilon(t)$ [right panel of Fig.3(a)]. A different behavior is observed in Fig.3(b), where the skin edge state cannot restore its original shape and $\epsilon(t)$ does not decay toward zero. We checked  \cite{Supplemental} that the self-healing property can be observed also when there are Bloch points (the GBZ zone crosses the unit circle) \rev{and for different types of scattering potentials, including inhomogeneous Hermitian and non-Hermitian amplifying potentials}.\\
 \\
\begin{figure}[t]
  \centering
    \includegraphics[width=0.45\textwidth]{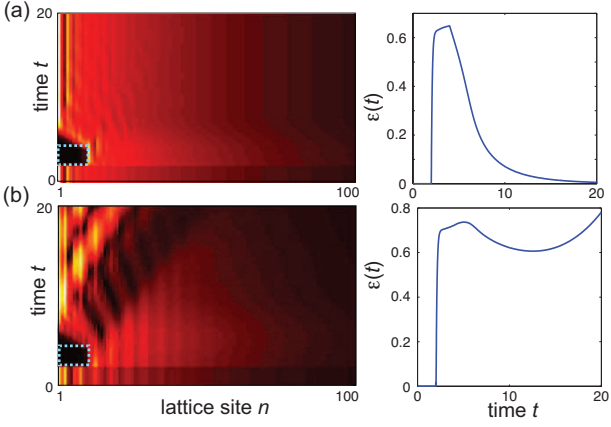}
   \caption{Self-healing of topological skin edge states. The left panels show the temporal evolution of the modulus of the normalized amplitudes $\tilde{\psi}_n(t)$  on a pseudo color map, in a semi-infinite lattice with parameter values as in Fig.2 and with an absorbing scattering potential (obstacle) localized in the dotted rectangular region of the space-time plane ($n \leq 10$ and $2<t<4$). The initial state $\psi_n(0)$ is the skin edge mode with energy \rev{$E_0=0.35 i$} in (a), and $E_0=-1+0.05 i$ in (b). The right panels show the corresponding temporal evolution of the function $\epsilon(t)$, defined by Eq.(\ref{eq2}), which measures the deviation of the evolved wave function from the skin state.}
     \label{fig3}
\end{figure}
 {\em Multiband systems.}  The previous analysis has been focused to single band models, however the self-healing property of topological skin edge states can be extended to multiband systems.
 As an illustrative example, we consider a quasi 1D lattice  composed by two side-coupled Hatano-Nelson chains  \cite{Hatano} [Fig.4(a)], which displays the critical NH skin effect \cite{r36}. The Bloch Hamiltonian of the systems reads
 \begin{equation}
 H(k)= \sigma_0 d_0 +t_0 \sigma_x +[V+i(\delta_{b}-\delta_{a}) \sin k ] \sigma_z
 \end{equation} 
where $d_0=2 t_1 \cos k -i(\delta_{a}+\delta_{b}) \sin k$,  $\sigma_l$ are the Pauli matrices, $(t_1 \pm \delta_{a,b})$ are the asymmetric left/right hopping amplitudes in the upper (a) and lower (b) chains, $ \pm V$ their on-site energy offset and $t_0$ is the side coupling constant. Figures 4(b,c) show a typical behavior of GBZ and energy spectra (PBC, OBC and SIBC) for $\delta_{a}>0$, $\delta_b<0$, with the shaded region corresponding to topological skin edge states localized at the left boundary under SIBC. Self-healing skin edge states are those with energy $E$ satisfying the condition ${\rm Im}(E)>E_m$, with $E_m={\rm max} (E_{m_1},E_{m_2})=E_{m1} \simeq 0.255$. The self-healing property is illustrated in Fig.4(d), where a skin edge state is scattered off by a complex absorbing potential in both chains ($V_n(t)=10 i$ for $4<t<8$ and $1 \leq n \leq 10$, $V_n=0$ otherwise).   
\begin{figure}[t]
  \centering
    \includegraphics[width=0.45\textwidth]{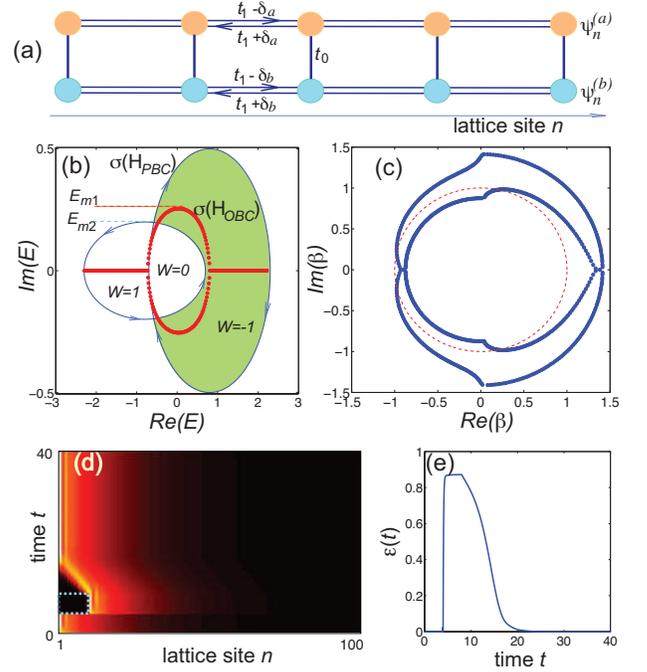}
   \caption{ (a) Scheme of two side-coupled Hatano-Nelson chains. (b) PBC (thin solid curves) OBC (solid dots) and SIBC (shaded area) energy spectra for $t_1=0.75$, $\delta_{a}=0.25$, $\delta_{b}=-0.15$, $t_0=0.05$ and $V=0.8$. The two PBC Bloch bands form two closed loops which are travelled in opposite direction, leading to three possible values $0, \pm 1$ of the winding $W$ in their interior. For any energy $E$ in the shaded area ($W=-1$) there is one topological edge state at the left boundary of the lattice. (c) Diagram of the GBZ (solid dots). The thin dashed curve shows the unit circle as a reference. (d,e) Self-healing of the topological edge state with energy $E=1+0.4i$. (d) Evolution of the normalized amplitudes $\sqrt{|\psi^{(a)}_{n}|^2+|\psi^{(b)}_{n}|^2} / \sum_n \sqrt{|\psi^{(a)}_{n}|^2+|\psi^{(b)}_{n}|^2}$, where $\psi_n^{(a)}$ and $\psi_n^{(b)}$ are the wave amplitudes at site $n$ in the two chains a and b, respectively. (e) Temporal behavior of $\epsilon(t)$. The absorbing scattering potential is localized in the dotted rectangular region of the space-time plane. }
     \label{fig3}
\end{figure}

\textit{Conclusion.} In summary, we have demonstrated that infinitely-many topological edge skin modes in semi-infinite NH lattices can exhibit self-healing properties, i.e. they can reconstruct their shape after being scattered off by a rather arbitrary space-time potential. Contrary to self-healing waves known in Hermitian systems, such as Bessel and Airy waves, the topological skin edge states are truly normalizable eigenstates of the underlying Hamiltonian. Our results unravel a fascinating fundamental property of recently-discovered topological skin modes, extend the idea of self-healing waves beyond the diffraction-free paradigm of Hermitian physics, and could be thus of potential relevance in different areas of physics and for future applications of self-healing NH waves.\\
\\
The author acknowledges the Spanish State Research Agency, through the Severo Ochoa
and Maria de Maeztu Program for Centers and Units of Excellence in R\&D (Grant No. MDM-2017-0711).


\begin{thebibliography}{99}


\bibitem{A1}
 Z. Bouchal, J. Wagner and M. Chlup, 
 Self-reconstruction of a distorted nondiffracting beam, Opt. Commun. {\bf 151}, 207 (1998).
\bibitem{A2}
D. McGloin and K. Dholakia, Bessel beams: Diffraction in a new light, Contemp. Phys. {\bf 46}, 15 (2005).
\bibitem{A3}
J. Broky, G.A. Siviloglou, A. Dogariu, and D.N. Christodoulides, 
Self-healing properties of optical Airy beams, Opt. Express {\bf 16}, 12880 (2008).
\bibitem{A4}
J. Durnin, J. J. Miceli, Jr., and J. H. Eberly, Diffraction-free beams,
Phys. Rev. Lett. {\bf 58}, 1499 (1987).
\bibitem{A5}
M.V. Berry and N. L. Balazs, Nonspreading wave packets,
Am. J. Phys. {\bf 47}, 264 (1979).
\bibitem{A6}
G. A. Siviloglou, J. Broky, A. Dogariu, and D. N. Christodoulides, Observation of Accelerating Airy Beams, Phys. Rev. Lett. {\bf 99}, 213901 (2007).
\bibitem{A8}
T. Ellenbogen, N. Voloch-Bloch, G.-P. Ayelet, and A. Arie, Nonlinear
generation and manipulation of Airy beams, Nat. Photon. {\bf 3}, 395
(2009).
\bibitem{A8a}
J. Lin, J. Dellinger, P. Genevet, B. Cluzel, F. de Fornel, and F. Capasso, Cosine-Gauss Plasmon Beam: A Localized Long-Range Nondiffracting Surface Wave,
Phys. Rev. Lett. {\bf 109}, 093904  (2012).
\bibitem{A8b}
R. Wang, Y. Wang, D. Zhang, G. Si, L. Zhu, L. Du, S. Kou, R. Badugu, M. Rosenfeld, J. Lin, P. Wang, H Ming, X. Yuan, and J.R.
Lakowicz, Diffraction-Free Bloch Surface Waves, ACS Nano {\bf 11}, 5383 (2017).
\bibitem{A8c}
M.S. Kim, A. Vetter, C. Rockstuhl, B.V. Lahijani, M. H\"ayrinen, M. Kuittinen, M. Roussey, and H.P. Herzig, 
 Multiple self-healing Bloch surface wave beams generated by a two-dimensional fraxicon, Commun Phys {\bf 1}, 63 (2018). 
\bibitem{A9}
P. Zhang, T. Li, J. Zhu, X. Zhu, S. Yang, Y. Wang, X. Yin, and X. Zhang,
Generation of acoustic self-bending and bottle beams by phase engineering,
Nat. Commun. {\bf 5}, 4316 (2014).
\bibitem{A10}
S. Fu, Y. Tsur, J. Zhou, L. Shemer, and A. Arie, Propagation dynamics
of Airy water-wave pulses, Phys. Rev. Lett. {\bf 115}, 034501 (2015).
\bibitem{A11}
Z. Lin, X. Guo, J. Tu, Q. Ma, J. Wu, and D. Zhang, Acoustic non-diffracting Airy beam,
J. Appl. Phys. {\bf 117}, 104503 (2015).
\bibitem{A12}
 G. Antonacci, D. Caprini, and G. Ruocco,
Demonstration of self-healing and scattering resilience of acoustic Bessel beams,
Appl. Phys. Lett. {\bf 114}, 013502 (2019).
\bibitem{A13}
N. Voloch-Bloch, Y. Lereah, Y. Lilach, A. Gover, and A. Arie, Generation of electron Airy beams,
Nature {\bf 494}, 331 (2013).
\bibitem{A14}
 V. Grillo, E. Karimi, G.C. Gazzadi, S. Frabboni, M.R. Dennis, and R.W. Boyd,
 Generation of Nondiffracting Electron Bessel Beams, Phys. Rev. X {\bf 4}, 011013 (2014).
\bibitem{A15}
 F. O. Fahrbach, P. Simon, and A. Rohrbach,  Microscopy with self-reconstructing beams, Nat. Photonics {\bf 4}, 780 (2010). 
\bibitem{A16} 
 T. A. Planchon, L. Gao, D. E. Milkie, M. W. Davidson, J.A. Galbraith, C.G Galbraith, and E. Betzig, Rapid three-dimensional isotropic imaging of living cells using Bessel beam plane illumination, Nat. Methods {\bf 8}, 417 (2011). 
\bibitem{A17}
S. Jia, J. C. Vaughan, and X. Zhuang, Isotropic three-dimensional
super-resolution imaging with a self-bending point spread function,
Nat. Photonics {\bf 8}, 302 (2014).
\bibitem{A18}
M. Duocastella and C. B. Arnold, Bessel and annular beams for materials processing , Laser Photonics Rev. {\bf 15}, 607 (2012)
\bibitem{A19}
V. Garces-Chavez, D. McGloin, H. Melville, W. Sibbett, and K. Dholakia,
Simultaneous micromanipulation in multiple planes using a self-reconstructing light beam, Nature {\bf 419}, 145  (2002).
\bibitem{A20}
J. Baumgartl, M. Mazilu, and K. Dholakia, Optically mediated particle
clearing using Airy wavepackets, Nat. Photonics {\bf 2}, 675 (2008).
\bibitem{A21}
M. McLaren, T. Mhlanga, M.J. Padgett, F.S. Roux, and A. Forbes, 
Self-healing of quantum entanglement after an obstruction, Nat. Commun. {\bf 5}, 3248 (2014).





\bibitem{r1}
Y. Ashida, Z. Gong, and M. Ueda,
Non-Hermitian Physics, Advances in Physics {\bf 69}, 3 (2020).

 
\bibitem{r1a}
T. E. Lee, Anomalous Edge State in a Non-Hermitian Lattice, Phys. Rev. Lett. {\bf 116}, 133903 (2016).
\bibitem{r2}
D. Leykam, K. Y. Bliokh, C. Huang, Y. D. Chong, and F.
Nori, Edge Modes, Degeneracies, and Topological Numbers in Non-Hermitian Systems, Phys. Rev. Lett. {\bf 118}, 040401 (2017).

\bibitem{r3}
Z. Gong, Y. Ashida, K. Kawabata, K. Takasan, S. Higashikawa, and M. Ueda, Topological Phases of Non-Hermitian Systems, Phys. Rev. X {\bf 8}, 031079 (2018).


\bibitem{r4}
H. Shen, B. Zhen, and L. Fu, Topological Band Theory for Non-Hermitian Hamiltonians, Phys. Rev. Lett. {\bf 120}, 146402 (2018).
\bibitem{r5}
F.K. Kunst, E. Edvardsson, J.C. Budich, and E.J. Bergholtz, Biorthogonal Bulk-Boundary Correspondence in Non-Hermitian Systems, Phys. Rev. Lett. {\bf 121}, 026808 (2018).


\bibitem{r6}
S. Yao and Z. Wang, Edge States and Topological Invariants of Non-Hermitian Systems, Phys. Rev. Lett. {\bf 121}, 086803 (2018).

\bibitem{r7}
V. M. Martinez Alvarez, J. E. Barrios Vargas, and L. E. F. Foa Torres, Non-Hermitian robust edge states in one dimension: Anomalous localization and eigenspace condensation at exceptional points, 
Phys. Rev. B {\bf 97}, 121401(R) (2018).
\bibitem{r8}
S. Yao, F. Song, and Z.Wang, Non-Hermitian Chern Bands, Phys. Rev. Lett. {\bf 121}, 136802 (2018).
\bibitem{r9}
C.H. Lee and R. Thomale, Anatomy of skin modes and topology in non-Hermitian systems, Phys. Rev. B {\bf 99}, 201103(R) (2019).
\bibitem{r10}
H. Zhou ‡ and J.Y. Lee, Periodic Table for Topological Bands with Non-Hermitian Bernard-LeClair Symmetries, Phys. Rev. B {\bf 99}, 235112  (2019).
\bibitem{r11}
C.-H. Liu, H. Jiang, and S. Chen, Topological classification of non-Hermitian systems with reflection symmetry,
Phys. Rev. B {\bf 99}, 125103 (2019).
\bibitem{r12}
C. H. Lee, L. Li, and J. Gong, Hybrid Higher-Order Skin-
Topological Modes in Non-reciprocal Systems, Phys. Rev. Lett.
{\bf 123}, 016805 (2019).
\bibitem{r13}
E. Edvardsson, F.K. Kunst, and E.J. Bergholtz, Non-Hermitian extensions of higher-order topological phases and their biorthogonal bulk-boundary correspondence,
Phys. Rev. B {\bf 99}, 081302(R) (2019).


\bibitem{r14}
F. Song, S. Yao, and Z. Wang, Non-Hermitian Topological Invariants in Real Space, Phys. Rev. Lett. {\bf 123}, 246801 (2019).


\bibitem{r15}
F. Song, S. Yao, and Z. Wang, Non-Hermitian skin effect and chiral damping in open quantum systems, Phys. Rev. Lett. {\bf 123}, 170401 (2019).


\bibitem{r16}
K. Yokomizo and S. Murakami, Non-Bloch Band Theory for Non-Hermitian Systems, Phys. Rev. Lett. {\bf 123}, 066404 (2019).


\bibitem{r17}
K. L. Zhang, H. C. Wu, L. Jin, and Z. Song, Topological phase transition independent of system non-Hermiticity,
Phys. Rev. B {\bf 100}, 045141 (2019).
\bibitem{r18}
L. Jin and Z. Song, Bulk-Boundary Correspondence in Non-Hermitian Systems in one dimension with chiral inversion symmetry, Phys. Rev. B {\bf 99}, 081103(R) (2019).
\bibitem{r19}
T. Liu, Y.-R. Zhang, Q. Ai, Z. Gong,
K. Kawabata, M. Ueda, and F. Nori, Second-order topological phases in non-Hermitian systems,
Phys. Rev. Lett. {\bf 122}, 076801 (2019).

\bibitem{r20}
A. Ghatak and T. Das, New topological invariants in non-Hermitian systems, J. Phys.: Condens. Matter {\bf 31}, 263001 (2019).

\bibitem{r20a}
K. Kawabata, K. Shiozaki, M. Ueda, and M. Sato, Symmetry and Topology in Non-Hermitian Physics, Phys. Rev. X 9, 041015 (2019).



\bibitem{r21}
L. Herviou, J.H. Bardarson, and N. Regnault,
Defining a bulk-edge correspondence for non-Hermitian Hamiltonians via singular-value decomposition,
Phys. Rev. A {\bf 99}, 052118 (2019).
\bibitem{r22}
S. Longhi, Probing non-Hermitian skin effect and non-Bloch phase transitions,
Phys. Rev. Research {\bf 1}, 023013 (2019).
\bibitem{r23}
D.S. Borgnia, A.J. Kruchkov, and R.-J. Slager, Non-Hermitian Boundary Modes and Topology, Phys. Rev. Lett. {\bf 124}, 056802  (2020).
\bibitem{r24}
 S. Longhi, Non-Bloch PT symmetry breaking in non-Hermitian photonic quantum walks,
Opt. Lett. {\bf 44}, 5804 (2019).
\bibitem{r25}
F.K. Kunst and V. Dwivedi, Non-Hermitian systems and topology: A transfer matrix perspective, Phys. Rev. B {\bf 99}, 245116 (2019).
\bibitem{r26}
K.-I. Imura and Y. Takane, Generalized bulk-edge correspondence for non-Hermitian topological systems,
Phys. Rev. B {\bf 100}, 165430 (2019).
\bibitem{r27}
N. Okuma and M. Sato, Topological Phase Transition Driven by Infinitesimal Instability: Majorana Fermions in Non-Hermitian Spintronics,
Phys. Rev. Lett. {\bf 123}, 097701 (2019).
\bibitem{r28}
J. Y. Lee, J. Ahn, H. Zhou, and A. Vishwanath, Topological Correspondence between Hermitian and Non-Hermitian Systems: Anomalous Dynamics,
Phys. Rev. Lett. {\bf 123}, 206404 (2019).
\bibitem{r28a}
S. Longhi, Topological phase transition in non-Hermitian quasicrystals, Phys. Rev. Lett. {\bf 122}, 237601 (2019).


\bibitem{r29}
N. Okuma, K. Kawabata, K. Shiozaki, and M. Sato, Topological Origin of Non-Hermitian Skin Effects,
Phys. Rev. Lett. {\bf 124}, 086801 (2020).

\bibitem{r30}
X. Zhang and J. Gong, Non-Hermitian Floquet topological phases: Exceptional points, coalescent edge modes, and the skin effect,
Phys. Rev. B {\bf 101}, 045415 (2020). 
\bibitem{r31}
K. Kawabata, N. Okuma, and M. Sato, 
Non-Bloch band theory of non-Hermitian Hamiltonians in the symplectic class,
Phys. Rev. B {\bf 101}, 195147 (2020).
\bibitem{r32}
S. Longhi, Non-Bloch-Band Collapse and Chiral Zener Tunneling,
Phys. Rev. Lett. {\bf 124}, 066602 (2020).

\bibitem{r33}
Z. Yang, K. Zhang, C. Fang, and J. Hu, Non-Hermitian Bulk-Boundary Correspondence and Auxiliary Generalized Brillouin Zone Theory, Phys. Rev. Lett. {\bf 125}, 226402 (2020).

\bibitem{r34}
K. Zhang, Z. Yang, and C. Fang,
Correspondence between winding numbers and skin modes in non-hermitian systems, Phys. Rev. Lett. {\bf 125}, 126402 (2020).

\bibitem{r34a}
Y. Yi and Z. Yang, Non-Hermitian Skin Modes Induced by On-Site Dissipations and Chiral Tunneling Effect,
Phys. Rev. Lett. {\bf 125}, 186802 (2020).
\bibitem{r34b}
N. Matsumoto, K. Kawabata, Y. Ashida, S. Furukawa, and M. Ueda, 
Continuous Phase Transition without Gap Closing in Non-Hermitian Quantum Many-Body Systems,
Phys. Rev. Lett. {\bf 125}, 260601 (2020).
\bibitem{r35}
C.H. Lee and S. Longhi, Ultrafast and anharmonic Rabi oscillations between non-Bloch bands, Commun. Phys. {\bf 3}, 147 (2020).

\bibitem{r36}
L. Li, Ching H. Lee, S. Mu, and J. Gong, Critical non-Hermitian Skin Effect, Nature Commun. {\bf 11}, 5491 (2020).


\bibitem{r37}
K. Kawabata, M. Sato, and K. Shiozaki,
Higher-order non-Hermitian skin effect,  Phys. Rev. B 102, 205118 (2020).
\bibitem{r38}
Y. Fu and S. Wan, Non-Hermitian Second-Order Skin and Topological Modes,  Phys. Rev. B {\bf 102}, 241202(R) (2020).
\bibitem{r39}
R. Okugawa, R. Takahashi, and K. Yokomizo, Second-order topological non-Hermitian skin effects, 
Phys. Rev. B {\bf 102}, 241202 (2020).
\bibitem{r40}
S. Longhi, Unraveling the non-Hermitian skin effect in dissipative systems,
Phys. Rev. B {\bf 102}, 201103 (2020).
\bibitem{r41}
C.-H. Liu, K. Zhang, Z. Yang, and S. Chen, Helical damping and dynamical critical skin effect in open quantum systems,
Phys. Rev. Research {\bf 2}, 043167 (2020).
\bibitem{r42}
E. Edvardsson, F.K. Kunst, T. Yoshida, and E.J. Bergholtz, Phase transitions and generalized biorthogonal polarization in non-Hermitian systems,
Phys. Rev. Res. {\bf 2}, 043046 (2020).
\bibitem{r43}
P. Gao, M. Willatzen, and J. Christensen,
Anomalous Topological Edge States in Non-Hermitian Piezophononic Media, Phys. Rev. Lett. {\bf 125}, 206402 (2020).
\bibitem{r44}
 L. Xiao, T. Deng, K. Wang, G. Zhu, Z. Wang, W. Yi, and P. Xue,
Observation of non-Hermitian bulk-boundary correspondence in quantum dynamics, Nature Phys. {\bf 16}, 761 (2020).
\bibitem{r45}
A. Ghatak, M. Brandenbourger, J. van Wezel, and C. Coulais, Observation of non-Hermitian topology and its bulk-edge correspondence, Proc Nat. Acad. Sci.  {`bf 117}, 29561 (2020).
\bibitem{r46}
T. Helbig, T. Hofmann, S. Imhof, M. Abdelghany, T. Kiessling, L.W. Molenkamp, C. H. Lee, A. Szameit, M. Greiter, and R. Thomale,  Generalized bulk-boundary correspondence in non-Hermitian topolectrical circuits, Nature Phys. {\bf 16}, 747 (2020).
\bibitem{r47}
T. Hofmann, T. Helbig, F. Schindler, N. Salgo, M. Brzezinska, M. Greiter, T. Kiessling, D. Wolf, A. Vollhardt, A. Kaba¨i, C.H. Lee, A. Bilusic, R. Thomale, and T. Neupert, Reciprocal skin effect and its realization in a topolectrical circuit, Phys. Rev. Research {\bf 2}, 023265 (2020).
\bibitem{r48}
S. Weidemann, M. Kremer, T. Helbig, T. Hofmann, A. Stegmaier, M. Greiter, R. Thomale, and A. Szameit, Topological funneling of light, Science {\bf 368}, 311 (2020).
\bibitem{r49}
Y. Song, W. Liu, L. Zheng, Y. Zhang, B. Wang, and P. Lu, Two-dimensional non-Hermitian Skin Effect in a Synthetic Photonic Lattice,
Phys. Rev. Applied {\bf 14}, 064076  (2020).
\bibitem{r50}
L. E. F. Foa Torres, Perspective on topological states of non-Hermitian lattices, 
J. Phys.: Materials {\bf 3}, 014002 (2020).

\bibitem{r51}
E.J. Bergholtz, J.C. Budich, and F.K. Kunst, Exceptional Topology in non-Hermitian Systems, Rev. Mod. Phys. {\bf 93}, 015005 (2021).

\bibitem{r52}
H. Hu and E. Zhao, 
Knots and Non-Hermitian Bloch Bands,
Phys. Rev. Lett. {\bf 126}, 010401 (2021).
\bibitem{r53}
K. Yokomizo and S. Murakami, Non-Bloch band theory in bosonic Bogoliubov-de Gennes systems, Phys. Rev. B {\bf 103}, 165123 (2021).
\bibitem{r54}
N. Okuma and M. Sato, Quantum anomaly, non-Hermitian skin effects, and entanglement entropy in open systems, Phys. Rev. B {\bf 103}, 085428 (2021).
\bibitem{r55}
H.-G. Zirnstein, G. Refael, and B. Rosenow, Bulk-Boundary Correspondence for Non-Hermitian Hamiltonians via Green Functions,
Phys. Rev. Lett. {\bf 126}, 216407 (2021).
\bibitem{r56}
T. Haga, M. Nakagawa, R. Hamazaki, and M. Ueda, Liouvillian Skin Effect: Slowing Down of Relaxation Processes without Gap Closing,
Phys. Rev. Lett. {\bf 127}, 070402 (2021).
\bibitem{r57}
J. Claes and T.L. Hughes, Skin effect and winding number in disordered non-Hermitian systems,
Phys. Rev. B {\bf 103}, L140201 (2021).


\bibitem{r58}
K. Wang, A. Dutt, K. Y. Yang, C.C. Wojcik, J. Vuckovic, and S. Fan, Generating arbitrary topological windings of a non- Hermitian band, Science {\bf 371}, 1240 (2021).
\bibitem{r59}
R. Okugawa, R. Takahashi, and K. Yokomizo, Non-Hermitian band topology with generalized inversion symmetry,
Phys. Rev. B {\bf 103}, 205205 (2021).
\bibitem{r60}
Z. Lin, L. Ding, S. Ke, and X. Li,
Steering non-Hermitian skin modes by synthetic gauge fields in optical ring resonators, Opt. Lett. {\bf 46}, 3512 (2021).
\bibitem{r61}
S. Longhi, Non-Hermitian topological phase transitions in superlattices and the optical Dirac equation, Opt. Lett. {\bf 46}, 4470 (2021).
\bibitem{r62}
N. Okuma and M. Sato,
Non-Hermitian Skin Effects in Hermitian Correlated or Disordered Systems: Quantities Sensitive or Insensitive to Boundary Effects and Pseudo-Quantum-Number, 
Phys. Rev. Lett. {\bf 126}, 176601 (2021).
\bibitem{r63}
C.-X. Guo, C.-H. Liu, X.-M. Zhao, Y. Liu, and S. Chen, Exact Solution of Non-Hermitian Systems with Generalized Boundary Conditions: Size-Dependent Boundary Effect and Fragility of the Skin Effect,
Phys. Rev. Lett. {\bf 127}, 116801 (2021).
\bibitem{r64}
K. Yokomizo and S. Murakami, Scaling rule for the critical non-Hermitian skin effect,
Phys. Rev. B {\bf 104}, 165117 (2021).
\bibitem{r65}
S. Longhi, Non-Hermitian skin effect beyond the tight-binding models,
Phys. Rev. B {\bf 104}, 125109 (2021).
\bibitem{r66}
K. Wang, A. Dutt, C.C. Wojcik, and S. Fan, Topological complex-energy braiding of non-Hermitian bands, 
Nature {\bf 598}, 59 (2021).
\bibitem{r67}
K. Wang, T. Li, L. Xiao, Y. Han, W. Yi, and P. Xue,
Detecting non-Bloch topological invariants in quantum dynamics, Phys. Rev. Lett. {\bf 127}, 270602 (2021) (2021).
\bibitem{r68}
K. Zhang, Z. Yang, and C. Fang,
Universal non-Hermitian skin effect in two and higher dimensions, arXiv:2102.05059 (2021).
\bibitem{r69}
D. Wu, J. Xie, Y. Zhou,  and J. An, 
Connections between the Open-boundary Spectrum and Generalized Brillouin Zone in
Non-Hermitian Systems, arXiv:2110.09259 (2021). 
\bibitem{r70}
W.-T. Xue, Y.-M. Hu, F. Song, and Z. Wang, 
Edge Burst in Non-Hermitian Quantum Walk, arXiv:2109.14628 (2021).

\bibitem{Referee1}
\rev{S. Weidemann, M. Kremer, S. Longhi, and A. Szameit, Topological triple phase transition in non-Hermitian Floquet quasicrystals, Nature {\bf 601}, 354 (2022).}
\bibitem{Referee2}
\rev{ Q. Lin, T. Li, L.  Xiao, K. Wang, W. Yi, and P. Xue, Simulating non-Hermitian quasicrystals with single-photon quantum walks, arXiv:2112.15024v1 (2022).}


\bibitem{Morandotti}
P. Zhang, Y. Hu, T. Li, D. Cannan, X. Yin, R. Morandotti, Z. Chen, and X. Zhang, Nonparaxial Mathieu and Weber Accelerating Beams,
Phys. Rev. Lett. {\bf 109}, 193901 (2012).

\bibitem{Yuce}
C. Yuce and Z.Turker, Self-acceleration in non-Hermitian systems, Phys. Lett. A {\bf 381}, 2235 (2017).



\bibitem{Supplemental}
See the Supplemental Material for (i) a brief review on energy spectra and GBZ in NH lattices, (ii) technical details on the proof of the theorem stated in the main text, (iii) further examples of self-healing of topological skin modes, and (iv) a method to generate skin edge states.

\bibitem{S1}
A. Calderon, F. Spitzer, and H. Widom, Inversion of Toeplitz matrices, Illinois J. Math. {\bf 3}, 490 (1959).
\bibitem{S2}
P. Schmidt and F. Spitzer, The Toeplitz matrices of an arbitrary Laurent polynomial, Math. Scand. {\bf 8}, 15 (1960).
\bibitem{S3}
I.I Hirschman, The spectra of certain Toeplitz matrices, Illinois J. Math. {\bf 11}, 145 (1967).
\bibitem{S4}
 A. B\"ottcher and S. M. Grudsky, {\it Spectral Properties of Banded Toeplitz Matrices} (SIAM, Philadelphia, 2005).
\bibitem{S5}
A. Iserles, How large is the exponential of a banded matrix?, New Zeland J. Math. {\bf 29}, 177 (2000).

\bibitem{rLonghi21}
\rev{S. Longhi, Selective and tunable excitation of topological non-Hermitian skin modes, arXiv:2112.04988 (2021).}

\bibitem{Hatano}
N. Hatano and D.R. Nelson, Localization Transitions in Non-Hermitian Quantum Mechanics, Phys. Rev. Lett. {\bf 77}, 570 (1996).




\end{thebibliography}
\end{document}